\documentclass[a4paper,12pt]{article}

\usepackage{amsmath}\usepackage{amssymb}\usepackage[dvips]{graphicx}
\usepackage{latexsym}\usepackage{cite}

\newcommand{\be}{\begin{equation}}
\newcommand{\ee}{\end{equation}}

\newcommand{\ka}{\kappa}
\newcommand{\mcA}{{\mathcal A}}
\newcommand{\mcD}{{\mathcal D}}

\newcommand{\mcN}{{\mathcal N}}

\def\beq{\begin{equation}}
\def\eeq{\end{equation}}
\def\pl{\partial}

\def\al{\alpha}
\def\bt{\beta}

\def\ga{\gamma}
\def\de{\delta}

\def\ka{\kappa}
\def\si{\sigma}
\def\Si{\Sigma}
\def\te{\theta}
\def\Te{\Theta}

\def\Om{\Omega}

\def\ep{\epsilon}
\def\ze{\zeta}

\def\sq{\sqrt}

\def\l{\left (}
\def\r{\right )}

\def\fr{\frac}
\def\la{\label}
\def\hs{\hspace}
\def\vs{\vspace}

\def\ran{\rangle}
\def\lan{\langle}
\def\ov{\overline}

\def\tm{\times}

\begin{document}

\begin{flushright}
CERN-PH-TH-2004-261\\
CFP-05-03\\
HD-THEP-04-54
\end{flushright}
\vspace{0.6cm}
\begin{center}
{\Large \bf Gauge and Modulus Inflation From 5D Orbifold SUGRA} 
\end{center}
\vspace{0.5cm}

\begin{center}
{\large   
Filipe Paccetti Correia$^a$\footnote{E-mail address: 
paccetti@fc.up.pt},
Michael G. Schmidt$^b$\footnote{E-mail address: 
M.G.Schmidt@thphys.uni-heidelberg.de},
Zurab Tavartkiladze$^c$\footnote{E-mail address: 
Zurab.Tavartkiladze@cern.ch}} 

\vspace{0.3cm}

$^a${\em 
Centro de F\' isica do Porto,
Faculdade de Ci\^ encias da Universidade do Porto\\
Rua do Campo Alegre 687, 4169-007 Porto, Portugal\\

$^b$
Institut f\"ur Theoretische Physik,
Universit\"at Heidelberg\\ 
Philosophenweg 16, 69120 Heidelberg, Germany\\

$^c$ Physics Department, Theory Division, CERN, CH-1211 Geneva 23, Switzerland
}
\end{center}
\vspace{0.4cm}
\begin{abstract}

We study the inflationary scenarios driven by a Wilson line field - 
the fifth component of 
a 5D gauge field and corresponding modulus field, within $S^{(1)}/Z_2$ orbifold supergravity (SUGRA). We use 
our off shell superfield formulation and give a detailed description of the 
issue of SUSY breaking by the $F$-component of the radion superfield. 
By a suitably gauged $U(1)_R$ symmetry and including couplings with 
compensator supermultiplets and a linear multiplet, we achieve a 
self consistent radion mediated SUSY breaking of no scale type. 
The inflaton 1-loop effective 
potential has attractive features needed for successful inflation.
An interesting feature of both presented inflationary scenarios are the 
red tilted spectra with $n_s\simeq 0.96$. For gauge inflation we obtain a 
significant tensor to scalar 
ratio ($r\approx 0.1$) of the density perturbations, while for the modulus inflation $r$ is strongly suppressed.

\end{abstract}

%
%


\vs{0.3cm}

\section{Introduction}

Inflation is the only candidate which naturally evades numerous cosmological 
problems \cite{Guth:1980zm}. In order to have a sufficiently flat universe, 
a de-Sitter type expansion with a slowly rolling scalar inflaton field  is
needed. This requires a flat inflaton potential and for that supersymmetry 
(SUSY) is believed to be crucial  in a realistic model building 
\cite{Copeland:1994vg}.
A different possibility for realizing a flat potential is that the inflaton 
field is a pseudo Nambu-Goldstone Boson (PNGB) field \cite{Freese:1990rb}. 
However, this idea seems to be difficult to realize since it usually requires 
VEVs much higher than the Planck scale: at such  large VEVs, one might not 
trust the results obtained in the framework of an (effective) quantum field 
theory (see a more detailed discussion in ref \cite{Arkani-Hamed:2003wu}). 
A nice and elegant realization of the PNGB inflation scenario was proposed in 
\cite{Arkani-Hamed:2003wu} (and subsequent works \cite{Kaplan:2003aj}, 
\cite{Hofmann:2003ag}), 
where an extra dimensional construction was suggested and the PNGB inflaton 
is the Wilson line field corresponding to the fifth component of a 5D $U(1)$ 
gauge field. In this setting, the flatness of the inflaton potential does not 
require unnatural assumptions and the model turns out to be fully self 
consistent with an effective 4D quantum field theory setting. 
Although the idea of ref. \cite{Arkani-Hamed:2003wu} works without invoking 
SUSY, we think that (with its phenomenological and theoretical motivations)  
it is worthwhile to study this type of scenarios in the framework of SUSY. Our 
recent work \cite{Hofmann:2003ag} was dedicated 
to this issue and can be considered as a step towards the construction 
of the SUSY gauge inflation scenario.
The setting which we have proposed there was based on a rigid on shell SUSY 5D 
construction of ref. \cite{Marti:2001iw} with the fifth dimension compactified 
on a  circle $S^{(1)}$. The radion superfield $T$ was used for SUSY 
breaking by its auxiliary component $F_T\neq 0$.

The aim of the present paper is to extend studies to 5D orbifold supergravity
(SUGRA). Our construction is based on the off shell formulation of 5D 
conformal SUGRA developed by Fujita, Kugo and Ohashi (FKO) 
\cite{Fujita:2001kv}, \cite{Fujita:2001bd}, \cite{Kugo:2002js}
and uses the superfield approach suggested recently by us in ref. 
\cite{PaccettiCorreia:2004ri} (see also the subsequent ref. 
\cite{Abe:2004ar})\footnote{For original papers on off shell 5D SUGRA formulation see refs \cite{zucker}. This formulation was used in many phenomenologically oriented papers \cite{phenor}.}. 
This superfield approach turned out to be  very economical and powerful 
for studying
various phenomenological and theoretical issues in 5D \cite{Correia:2004pz}. 
Here, we first present in section 2 the minimal setting which is 
needed in order to realize gauge inflation. Then in section 3 we 
give full account of the issue of SUSY breaking by the radion's auxiliary 
$F$-component. We show that to obtain flatness and selfconsistency, 
gauging of a $U(1)_R$ part of the global $SU(2)_R$ 
symmetry and a linear multiplet with appropriate couplings with the 
$U(1)_R$ gauge supermultiplet play an important role. In section 4 we turn 
to the calculation of the one loop inflaton effective potential 
including the Wilson line 
and modulus fields. We separately study two inflationary scenarios:
in section 5 the gauge inflation and in section 6 the inflation driven by the modulus field.
We discuss  features allowing to realize 
a natural inflation and give a detailed study of some properties for both 
inflationary scenarios.


\section{The setting}

In this letter we will deal with two types of hypermultiplets. 
One is a compensator (denoted by ${\bf h}$) and is necessary for gauge fixing 
of the conformal symmetry. The second type of hypermultiplet is a physical one-
referred to as a matter hypermultiplet (denoted by ${\bf H}$). It will play 
an important role in the generation of the inflaton (the Wilson line field) potential. In general, 5D hypermultiplets 
${\mathbb H}^{\al }=(\mcA^{\al }_i,~\ze^{\al }, {\cal F}^{\al }_i)$ 
can be ordered into $r$ pairs
$({\mathbb H}^{2\hat{\al }-1},~{\mathbb H}^{2\hat{\al }})$, 
where $\hat{\al }=1, 2, \cdots , r$ (see 
\cite{Fujita:2001kv}-\cite{Kugo:2002js} for a detailed discussion). 
In the following, we will use the notation
\beq
{\bf H}\equiv ({\bf H}_1,~{\bf H}_2)=(H, ~H^c)~,
\la{hyper}
\eeq
for such a pair (omitting the index $\hat{\al }$) and similar for the 
compensator ${\bf h}=({\bf h}_1, {\bf h}_2)=(h, h^c)$. In the general 
discussion of the hypermultiplet case we will use ${\bf H}$ and understand 
that this similarly applies to the compensator. Essential differences  for 
the compensator hypermultiplet, will be pointed out throughout the text. 
The 5D hypermultiplet of eq. (\ref{hyper}) decomposes into a 
pair of $\mcN=1$ 4D chiral superfields
with opposite orbifold parities \cite{Kugo:2002js}:
$$
H=\l {\cal A}^{2\hat{\al }}_2=({\cal A}_1^{2\hat{\al }-1})^*,~
-2{\rm i}\ze^{2\hat{\al }}_R,~
({\rm i}M_*{\cal A}+{\hat{\mcD}}_5{\cal A})^{2\hat{\al }}_1\r, 
$$
\beq
H^c=\l {\cal A}^{2\hat{\al }-1}_2=-({\cal A}^{2\hat{\al }}_1)^*,~
-2{\rm i}\ze^{2\hat{\al }-1}_R,~
({\rm i}M_*{\cal A}+{\hat {\mcD}}_5{\cal A})^{2\hat{\al }-1}_1\r, 
\la{hyperdec}
\eeq
with 
$$
M_*{\cal A}^{\al }_i=igM^I(t_I)^{\al }_{\bt }{\cal A}^{\bt }_i+
{\cal F}^{\al }_i~,
$$
\beq
\hat{\mcD}_{5 }{\cal A}^{\al }_i=
\pl_{5 }{\cal A}^{\al }_i-i
gW_{5 \bt }^{\al }{\cal A}^{\bt }_i-
W_{5 }^0\fr{1}{\al }{\cal F}^{\al }_i-
\ka V_{5 ij}{\cal A}^{\al j}-
2\ka {\rm i}\bar \psi_{5 i}\ze^{\al }~,
\la{covders}
\eeq 
where $(t_I)^{\al }_{\bt }$ is the generator of the gauge group
$G_I$. 
In this paper we will deal only with abelian gauge groups. 
In this case the gauge coupling $g$ should be replaced by $g/2$.
The components $(\phi ,\psi , F_{\Phi } )$ of a 4D chiral superfield $\Phi $ 
are assumed to be of 'right handed' chirality. 
Therefore, the superfield with a left chirality in a two component notation
is given by
\beq
\Phi =(\phi ,\psi , F_{\phi })=\phi^*+\te \psi_L-\te^2F_{\Phi }^*~.
\la{Phisup}
\eeq
We will use this basis during the calculations.

Besides the hypermultiplets, in this discussion, three 
5D gauge supermultiplets $(V, \Si )^I$ will be considered. 
{\bf i)} The $U(1)_Z$ central charge symmetry
corresponds to $I=0$: $(V, \Si )^{I=0}\equiv (V_0, \Si_0)$. 
This is a compensating gauge supermultiplet and, as was 
observed in \cite{PaccettiCorreia:2004ri}, in the rigid limit accounts for the radion superfield.
In the covariant derivatives of eq. (\ref{covders}) $I=0$ does not participate 
in $M^I(t_I)^{\al }_{\bt }$ and $W^{\al }_{5\bt }$, but acts only on an 
auxiliary component ${\cal F}^{\al }_i$. This is a particular property of 
the compensating $I=0$ gauge supermultiplet.
{\bf ii)} Our construction is based on a gauged $U(1)_R$ symmetry whose corresponding 
gauge supermultiplet is $(V, \Si )^{I=R}\equiv (V_R, \Si_R)$. 
Only the compensator hypermultiplet ${\bf h}$ is charged under this group.
{\bf iii)} Finally, we introduce an $U(1)$ gauge supermultiplet
$(V, \Si )^{I=1}\equiv (V_1, \Si_1)$, where $\Si_1$ contains the fifth 
component of a vector field which generates the Wilson line field 
$\Te =\int dyA_5$. The latter being a 4D scalar will play the role of the 
inflaton field in the following. Note that only the matter hypermultiplet 
${\bf H}$ is charged under $(V_1, \Si_1)$.

The orbifold $Z_2$ parities ($y\to -y$) of the introduced gauge supermultiplets  
are given as:
\beq
Z_2:~~~(V_0, V_R, V_1)\to -(V_0, V_R, V_1)~,~~~~~
(\Si_0, \Si_R, \Si_1)\to (\Si_0, \Si_R, \Si_1)~.
\la{paritVSi}
\eeq
Therefore all 4D gauged $U(1)$ symmetries are broken on the orbifold fixed 
points.
As far as the hypermultiplets are concerned,
without loss of generality we can consider the following orbifold parity
prescriptions:
\beq
Z_2:~~~H\to H~,~~~~~H^c\to -H^c~.
\la{1Hpar}
\eeq
{}For all gauge fields $(V_I, \Si_I)$ we introduce the following 
parameterization
\beq
{\bf V}^{ab}=gV \vec{q}\cdot \vec{\si }^a_b~,~~~~
{\bf \Si }^{ab}=g\Si \vec{q}\cdot \vec{\si }^a_b~,~~~
{\rm with}~~~|\vec{q}|=1~.
\la{paramVSi}
\eeq
With this, the hypermultiplet Lagrangian is given by \cite{Correia:2004pz}
\beq
e_{(4)}^{-1}{\cal L}({\rm hyper})= 
\int d^4\te
{\mathbb W}_y 2 {\bf H}_a^{\dagger }(e^{-2{\bf V}})^{ab}{\bf H}_b-
\int d^2\te 
({\bf H}\ep )_a(\hat{\pl }_y-{\bf \Si })^{ab}{\bf H}_b+{\rm h.c.} ~
\la{hypActneg}
\eeq
In case of the compensator,  the Lagrangian eq.\eqref{hypActneg}  should come with 
opposite sign.  
The superoperator $\hat{\pl }_y$ is obtained by promoting ${\pl }_y$
to an operator containing odd (under orbifold parity) elements of the 5D
Weyl multiplet (see \cite{PaccettiCorreia:2004ri} for a more detailed 
discussion), which do not have any relevance for our purposes and can be 
ignored. 

With the orbifold parity assignments given in (\ref{paritVSi}) 
and (\ref{1Hpar})
we should gauge the $U(1)$ symmetries of (\ref{paramVSi}) in  $\si^1, \si^2$
direction, i.e. $q_3=0$ \footnote{The other possibility would be to gauge in the $\sigma^3$-direction which implies the introduction of an odd gauge coupling. This was used to obtain supersymmetric Randall-Sundrum models (see \cite{PaccettiCorreia:2004ri} and references therein, also \cite{Flacke:2005tm} for a recent study). From couplings in the $\sigma^3$-direction we obtain effective potentials which are flat in the Wilson-line $\Theta$-direction. For this reason we don't consider such couplings in this work.}.
In (\ref{hypActneg}) ${\mathbb W}_y$ is a real general type  4D supermultiplet
which contains part of the radion chiral superfield as 
\cite{PaccettiCorreia:2004ri}
\beq
{\mathbb W}_y=\fr{1}{2}(T+T^{\dagger })+\cdots 
\la{TinW}
\eeq
This relation is useful to account for the radion coupling with 
hypermultiplets. Since all 4D gauge superfields  $V$ have negative orbifold 
parities in this setting, they 
do not contain zero mode states and will be irrelevant for us. Therefore, 
we will set further $V=0$. Taking all this into account, the action (\ref{hypActneg})
for matter and compensator hypermultiplets can be written as:
$$
{\cal L}({\rm hyper})|_{V=0}={\cal L}(H)+{\cal L}(h)~,
$$
$$
e_{(4)}^{-1}{\cal L}(H)=\hs{-0.2cm}\int d^4\te 
(T+T^{\dagger })\l 
H^{\dagger }H+H^{c\dagger }H^c \r +\hs{-0.2cm}
\int d^2\te \l 2H^c\pl_yH+g_1\Si_1(e^{{\rm i}\hat{\te }_1}H^2-
e^{-{\rm i}\hat{\te }_1}H^{c2})\r +{\rm h.c.}
$$
\beq
e_{(4)}^{-1}{\cal L}(h)=\hs{-0.1cm}-\hs{-0.1cm}\int \hs{-0.1cm}d^4\te 
(T+T^{\dagger })\l 
h^{\dagger }h+h^{c\dagger }h^c \r -\hs{-0.2cm}
\int \hs{-0.1cm}d^2\te \l 2h^c\pl_yh+g_R\Si_R(e^{{\rm i}\hat{\te }_R}h^2-
e^{-{\rm i}\hat{\te }_R}h^{c2})\r +{\rm h.c.}
\la{hypAct1}
\eeq
where $\cos \hat{\te }_1=q_1^1$, $\sin \hat{\te }_1=q_2^1$,
$\cos \hat{\te }_R=q_1^R$, $\sin \hat{\te }_R=q_2^R$.

\section{SUSY breaking through the radion superfield}

In our model, for SUSY breaking we will use a non-zero $F$ component 
of the radion superfield $T$. As it was pointed out in ref.\cite{PaccettiCorreia:2004ri}, to obtain a flat tree-level potential for $F_T$ we need to introduce a 
linear multiplet ${\bf L}$ which couples with the $V^{I=R}$ vector 
multiplet. As we will see below, the 
r\^ ole of ${\bf L}$ is to insure a self consistent SUSY breaking. 
Assuming that ${\bf L}$ is neutral under $V^{I=R}$, its field content
is \cite{Fujita:2001bd}
\beq
{\bf L}=(L^{ij}~,\phi^i~,E^{\mu \nu }~,N)~,
\la{Lcontent}
\eeq
where $E^{\mu \nu }$ is an unconstrained antisymmetric tensor field.
The coupling action of ${\bf L}-V^{I=R}$ is given by
$$
e^{-1}{\cal L}(V^{I=R}, {\bf L})= Y_R^{ij}L_{ij}+2\ov{\Om }_R^i\phi_i+
2{\rm i}\ov{\psi }_i^a\ga_a\Om_{Rj}L^{ij}+
$$
\beq
\fr{1}{2}M_R\l N-2{\rm i}\ov{\psi }_b\ga^b\phi -
2{\rm i}\ov{\psi }_a^i\ga^{ab}\psi^j_bL_{ij}\r +
\fr{1}{4}e^{-1}F_{\mu \nu }(W_R)E^{\mu \nu }~.
\la{LVR}
\eeq 
${\bf L}$ plays the role of a Lagrange multiplier. A variation with 
respect to the components of ${\bf L}$ leads to
\beq
M_R=0~,~~~~\ov{\Om}_R^i=0~,~~~~Y_R^{ij}=0~,~~~~F_{\mu \nu }(W_R)=0~.
\la{compVR}
\eeq
The last equation of (\ref{compVR})  has the solution
\beq
W_{5R}={\rm constant}~.
\la{solW5R}
\eeq
This is enough to insure a non zero $F$ component of $T$.

Consider the part of the action (\ref{hypAct1})  which involves the 
compensator hypermultiplet. The relevant bosonic couplings have the form
$$
e_{(4)}^{-1}{\cal L}(h)\supset -2\left |F_h+\pl_5 h^{c*}-
g_R\Si_R^*e^{{\rm i}\hat{\te }_R}h^*+\fr{1}{2}F_Th\right |^2+
2\left |\pl_5h^{c*}-g_R\Si_R^*e^{{\rm i}\hat{\te }_R}h^*+
\fr{1}{2}F_Th\right |^2-
$$
$$
2\left |F_{h^c}-\pl_5 h^*+
g_R\Si_R^*e^{-{\rm i}\hat{\te }_R}h^{c*}+\fr{1}{2}F_Th^c\right |^2+
2\left |\pl_5h^*-g_R\Si_R^*e^{-{\rm i}\hat{\te }_R}h^{c*}-
\fr{1}{2}F_Th^c\right |^2+
$$
\beq
g_RF_{\Si_R}^*\l e^{{\rm i}\hat{\te }_R}(h^*)^2-
e^{-{\rm i}\hat{\te }_R}(h^{c*})^2\r +{\rm h.c.}
\la{hbos}
\eeq
where for the lowest components of the hypermultiplet we have used the same 
notation as for the corresponding superfield.
From (\ref{hyperdec}), (\ref{covders}) we have 
$$
F_h={\cal F}_h-\pl_5h^{c*}-{\rm i}\fr{g_R}{2}W_{5R}e^{{\rm i}\hat{\te}_R}h^*+
\fr{1}{2}F_Th~,
$$
\beq
F_{h^c}={\cal F}_{h^c}+\pl_5h^*+
{\rm i}\fr{g_R}{2}W_{5R}e^{-{\rm i}\hat{\te}_R}h^{c*}+
\fr{1}{2}F_Th^c~,~~
{\rm with}~~{\cal F}_h=({\rm i}-\fr{W_5^0}{\al }){\cal F}^{2\hat{\al }}_{h1}~,
~~{\cal F}_{h^c}=({\rm i}-\fr{W_5^0}{\al }){\cal F}^{2\hat{\al }-1}_{h1}~,
\la{FRels}
\eeq
where the relations
\beq
\Si_R=-\fr{\rm i}{2}W_{5R}~,~~~~
\ka (V_5^1+{\rm i}V_5^2)=-\fr{{\rm i}F_T}{e_y^5}~,~~~{\rm with}~~~e_y^5=1
\la{FTSiRels}
\eeq
have been used. Taking into account all this and the constraints
$h=\ka^{-1}, h^c=0$, (\ref{hbos}) reduces to
$$
e_{(4)}^{-1}{\cal L}(h)\supset 
-2\left | {\cal F}_h-{\rm i}g_RW_{5R}e^{{\rm i}\hat{\te }_R}\ka^{-1}+
F_T\ka^{-1}\right |^2-2\left | {\cal F}_{h^c}\right |^2+
$$
\beq
\fr{\ka^{-2}}{2}\left | F_T-{\rm i}g_RW_{5R}e^{{\rm i}\hat{\te }_R}\right |^2
+g_R\ka^{-2}(F_{\Si_R}e^{-{\rm i}\hat{\te }_R}+{\rm h.c.})~.
\la{hbos1}
\eeq
The on-shell equations 
$\fr{\pl {\cal L}}{\pl {\cal F}_h}=\fr{\pl {\cal L}}{\pl {\cal F}_{h^c}}=
\fr{\pl {\cal L}}{\pl F_T}=0$ then have the solutions
\beq
F_T={\rm i}g_RW_{5R}e^{{\rm i}\hat{\te }_R}~,~~~~{\cal F}_h={\cal F}_{h^c}=0~.
\la{solFs}
\eeq
Therefore, gauging $U(1)_R$  we have obtained a non zero $F_T$ with a flat 
potential. This is a no-scale SUSY breaking scenario with SUSY 
breaking mediated by the radion superfield\footnote{In the FKO treatment the 
role of a $F_T$-VEV is played by the gauge field component's $W_{5R}$ VEV 
after a redefinition $F_T^N=F_T-{\rm i}g_RW_{5R}e^{{\rm i}\hat{\te }R}$ with
$\lan F_T^N\ran =0$.}. For a discussion of this 
phenomenon within a 5D on shell construction see \cite{luty1}, \cite{luty2}.
An $F_T\neq 0$ is important for transmitting the SUSY breaking into 
the matter sector. {\it All} states which carry an $SU(2)_R$ index, couple 
with $F_T$ through the covariant derivative and obtain a soft SUSY breaking 
mass. For instance, the zero mode of the 4D gravitino obtains a soft mass 
through the 5D gravitino kinetic term by mixing with 
$\psi_{5-}\equiv {\rm i}(\psi_{5L}^1+\psi_{5R}^2)$. The latter is a 
goldstino - the fermionic component of the radion superfield.

Concluding this section, let us comment on another role of the linear 
multiplet. It insures that all other $F$-terms are zero. The first term of 
(\ref{LVR}) can be written as
\beq
Y_R^{ij}L_{ij}=2Y_R^aL^a=(F_{\Si_R}L+{\rm h.c.})+\cdots ~,
\la{inFs}
\eeq
where $L=-L^1+{\rm i}L^2$ and ellipses stand for terms which are irrelevant 
for us. Collecting together all couplings containing $F_{\Si_R}$, we have
\beq
e_{(4)}^{-1}{\cal L}(F_{\Si_R})=2\left |F_{\Si_R}\right |^2+
F_{\Si_R}(g_R\ka^{-2}e^{-{\rm i}\hat{\te }_R}+L)+{\rm h.c.}
\la{allFSiR}
\eeq
The conditions 
$\fr{\pl {\cal L}}{\pl F_{\Si_R}}=\fr{\pl {\cal L}}{\pl L}=0$ are satisfied 
by the solutions
\beq
F_{\Si_R}=0~,~~~~~L=-g_R\ka^{-2}e^{-{\rm i}\hat{\te }_R}~.
\la{solFSiRL}
\eeq
Note, that without the coupling to the lowest component of the linear multiplet, we would not be 
able to have $F_{\Si_R}=0$. The latter is needed for the $F$-flatness  
and a vanishing vacuum energy on the classical level.

Together with the gauge inflation, below we will also study the inflation driven by the modulus field $M^1$. With $\lan M^1\ran \neq 0$ the corresponding 
$F_{\Si_1}$ will have the potential:\footnote{For the definition of 
prepotential $P({\cal V}_5)$ see \cite{PaccettiCorreia:2004ri}.}
$\int de^4\te 2{\mathbb W}_yP({\cal V}_5)\to 
-{\cal N}_{11}|F_{\Si_1}-\fr{1}{2}M^1F_T|^2$. This gives
\beq
F_{\Si_1}=\fr{1}{2}M^1F_T~,
\la{FSi1}
\eeq
which will play an important role for calculation of the masses of KK states.

\section{KK decomposition and inflaton potential}

Now we are ready to derive the inflation effective potential. 
Relevant for us is the 4D chiral superfield 
$\Si_1$ which contains the
fifth component $A_5^1$ of $U(1)$ and the corresponding (real) modulus $M^1$ as 
$\Si_1= \fr{1}{2}(M^1-{\rm i}A_5^1)$. The superfield
$\Si_1$ has positive $Z_2$ orbifold parity and therefore the Wilson line field
\beq
\Te =\int dy A_5=2\pi RA_5^1~,
\la{wilson}
\eeq
and (the zero mode of) $M^1$ are $y$-independent 4D scalars. Taking all this into account, from 
(\ref{hypAct1}) with (\ref{FSi1})
one can easily derive the potential for the scalar components $H , H^c$ with phase redefinition
$H\to e^{{\rm i}\hat{\te }_1/2}H$, $H^c\to e^{-{\rm i}\hat{\te }_1/2}H^c$:
$$
V(H)=2\left | \pl_5H-\fr{g_1}{2}(M^1-\fr{{\rm i}\Te }{2\pi R})H^c 
-\fr{1}{2}F_T^*H^{c*}\right |^2 
+2\left | \pl_5 H^c -\fr{g_1}{2}(M^1-\fr{{\rm i}\Te }{2\pi R})H
+\fr{1}{2}F_T^*H^*\right |^2+
$$
\beq
\fr{1}{2}g_1M^1F_T(H^2-H^{c2})+\fr{1}{2}g_1M^1F_T^*(H^{*2}-H^{c*2})~.
\la{Vphi}
\eeq 
With  the 
parity assignment (\ref{1Hpar}), the KK decomposition for $H $,
$H^c$ is given by
\beq 
H=\fr{1}{2\sq{\pi R}}H^{(0)}+
\fr{1}{\sq{2\pi R}}\sum_{n\neq 1}H^{(n)}\cos \fr{ny}{R}~,~~~
~~
H^c=\fr{1}{\sq{2\pi R}}
\sum_{n\neq 1}\ov{H}^{\hs{0.6mm}(n)}\sin \fr{ny}{R}~.
\la{decKKphi}
\eeq
Upon integration along the fifth dimension 
${\cal L}^{(4)}=\int_0^{2\pi R}dy {\cal L}^{(5)}$,
one can easily see that the mass$^2$ of the two real zero modes are
\beq
(m_{\pm }^{(0)})^2=\fr{1}{R^2}(\fr{g_1\Te }{4\pi }\pm \fr{1}{2}R|F_T|)^2~.
\la{zeroMphi}
\eeq
{}For KK states it is convenient to choose the basis 
$H^{(n)}=\fr{\rm i}{\sq{2}}(H^{(n)}_{-}-H^{(n)}_{+})$,
$\ov{H}^{(n)}=\fr{1}{\sq{2}}(H^{(n)}_{-}+H^{(n)}_{+})$.
The mass$^2$ matrices for appropriate $n$-th KK modes are 
\begin{equation}
\begin{array}{cc}
 & {\begin{array}{cc}
\hspace{-0.5cm} H^{(n)}_{-}\hspace{2.2cm} & 
\hspace{3cm}H^{\hs{0.6mm}(n)*}_{-} \hspace{0.1cm} 
\end{array}}\\ \vspace{5mm}
\begin{array}{c} 
H^{(n)*}_{-} \\ H^{\hs{0.6mm}(n)}_{-} 
 \end{array}\!\!\!\!\! \hs{-0.1cm}&{\left(\begin{array}{cccc}
(n\hs{-0.6mm}+\hs{-0.6mm}\fr{g_1\Te }{4\pi })^2\hs{-0.6mm}+\hs{-0.6mm}
\fr{1}{4}|RF_T|^2+\fr{1}{4}(g_1RM^1)^2 &
{\rm i}(n+\fr{g_1\Te }{4\pi })RF_T^* 
\vs{0.2cm}
\\  
\hs{-0.2cm} -{\rm i}(n+\fr{g_1\Te }{4\pi })RF_T^*  &
\hs{-0.1cm} 
(n\hs{-0.6mm}+\hs{-0.6mm}\fr{g_1\Te }{4\pi })^2\hs{-0.6mm}+\hs{-0.6mm}
\fr{1}{4}|RF_T|^2+\fr{1}{4}(g_1RM^1)^2 
\hs{-0.1cm}\end{array}\hs{-0.1cm}\right)\fr{1}{2R^2}}~,
\end{array}  \!\!  
\label{matrixphi} 
\end{equation} 
and similar for $H^{\hs{0.6mm}(n)}_{+}$ states.
For mass$^2$'s of four real scalar states (per $n\neq 0$ KK state) 
we thus get
\beq
[m^{(n)}(H_{-})]^2_{\pm }=[m^{(n)}(H_{+})]^2_{\pm }=
\fr{1}{R^2}(n+\fr{g_1\Te }{4\pi }\pm \fr{1}{2}|RF_T|)^2+
\fr{1}{2}(g_1M^1)^2~.
\la{phiKKm}
\eeq

A non-zero $F_T$ does not affect the masses of the fermionic components 
($\psi_H$, $\ov{\psi }_{H^c}$) coming from
$H_M, H_M^c$ because they are blind with respect of $SU(2)_R$. 
Therefore the masses of Majorana fermionic components are
$$
m^{(n)}(\psi_H)=\fr{1}{R}(n-g_1\fr{\Te }{4\pi })~,
~~~~n=-\infty ,\cdots ,\infty ~,
$$
\be
m^{(n)}(\ov{\psi }_{H^c})=
\fr{1}{R}(n+g_1\fr{\Te }{4\pi })~,~~~~n=-\infty,\cdots ,\infty ~,~~~n\neq 1~.
\la{psiKK}
\eeq
Note that the spectrum in (\ref{phiKKm}), (\ref{psiKK}) is equivalent to one obtained
within the Scherck-Schwarz SUSY breaking scenario.

As we have already mentioned, integration of the states with 
$\Te $-dependent masses induces an effective 1-loop potential for $\Te $. 
The potential will also depend on the modulus $M^1$.
Using
Poisson resummation for each KK mode's contribution to the effective 
potential  
(or starting from a worldline expression, see an appendix in 
\cite{Hofmann:2003ag}), we finally obtain
$$
{\cal V}^{\rm eff}(\phi_{\Te })=\fr{3}{16\pi^6R^4}
\sum_{k=1}^{\infty} \fr{1}{k^5}\l 1-\cos (\pi kR|F_T|)\r \cdot
\cos(\pi kg_4R\phi_{\Theta })\tm 
$$
\beq
e^{-\pi kg_4R|\phi_M|}\l 1+\pi kg_4R|\phi_M| +\fr{1}{3}(\pi kg_4R|\phi_M|)^2
\r ~.
\la{Vinfl}
\eeq 
The effective potential in (\ref{Vinfl}) is written in terms of canonically 
normalized 4D scalar
fields $\phi_{\Te }=\Te /\sq{2 \pi R}$, $\phi_{M}=\sq{2\pi R}M^1$
and dimensionless 4D gauge coupling $g_4=g_1/\sq{2\pi R}$ (for a general 
cubic norm function the field $\phi_M$ is canonically normalized in the global minimum with $\lan \phi_M\ran =0$).

Notice that the divergent bosonic and fermionic
contributions at $k=0$ cancel exactly because of SUSY.
In the limit $F_T\to 0$ (unbroken SUSY) the effective one-loop potential
vanishes.
The potential in (\ref{Vinfl}) is invariant under the shifts
$\phi_{\Te }\to \phi_{\Te }+\fr{2k_1}{kg_4R}$, $|F_T|\to |F_T|+\fr{2k_2}{kR}$
($k_{1,2}=$integer) reflecting the invariance under 5D gauge symmetries.
Besides the $\phi_{\Theta }$ ($\phi_{M}$)-dependent part, the potential 
gets a constant contribution by integration of states which are neutral under $(V_1, \Si_1)$ but feel SUSY breaking through $F_T$. These kind of 
states are for example the 4D gravitino, the gauginos and $(V_1, \Si_1)$ neutral bulk hypermultiplets. Thus we add a constant part to the potential in eq. (\ref{Vinfl}) 
and tune the former in such a way that the potential is zero in the global 
minimum (this is the usual fine tuning of the 4D cosmological constant). 
Keeping the dominant terms of (\ref{Vinfl}), the inflaton potential will have 
the form
\beq
{\cal V}={\cal V}^{\rm eff}\left. \right |_{k=1}+{\cal V}_0 ~,~~~~{\rm with}~~~~
{\cal V}_0=\fr{3}{16\pi^6R^4}\l 1-\cos (\pi R|F_T|)\r ~.
\la{Vinfl1}
\eeq 
It's profile is plotted in Fig. 1.
For $F_T\not=0$ and $g_4R\phi_{\Theta }\neq 1$, $\phi_M\neq 0$ the potential
${\cal V}$ is positive, and thus it drives de Sitter expansion. 
Since ${\cal V}$ depends on two dynamical fields we will have inflation 
driven by these two fields. Below we will study the two extreme cases where one of 
the fields lies in its minimum and the inflation is driven by only one field. 
This allows an analytical study of the inflation and spectral properties 
of the density perturbations. 
The analysis for two field inflation will be presented elsewhere.

\begin{figure}
\begin{center}
\resizebox{0.6\textwidth}{!}{
  \includegraphics{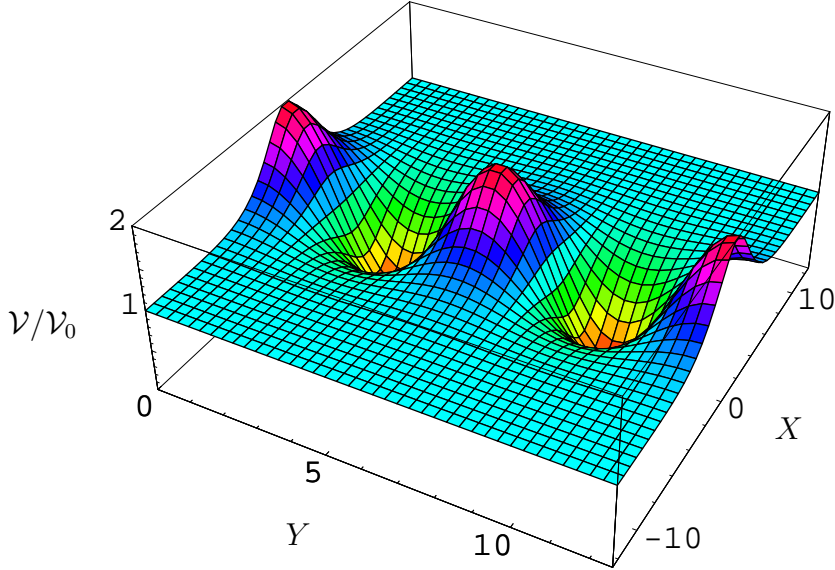}
}
\put(-215,20){$Y$}
\put(-30,60){$X$}
\put(-320,100){${\cal V}/{\cal V}_0$}
\caption{Effective potential as a function of $X=\pi g_4 R \phi_M$ and $Y=\pi g_4 R \phi_{\Theta}$.}
\label{fig:1}       
\end{center}
\end{figure}

\section{Gauge inflation}

First we consider the inflation driven by $\phi_{\Te }$ and set $\phi_M=0$.
In this case
the two 'slow roll' parameters are
$$
\ep =\fr{(M_{\rm Pl})^2}{2}\l \fr{{\cal V}'}{{\cal V}}\r^2 =\fr{\pi^2}{2}
(g_4RM_{\rm Pl})^2\tan^2\fr{\pi g_4R\phi_{\Theta }}{2}~,
$$
\beq
|\eta |=(M_{\rm Pl})^2\left |\fr{{\cal V}''}{{\cal V}}\right |=\fr{\pi^2}{2}
(g_4RM_{\rm Pl})^2|\tan^2\fr{\pi g_4R\phi_{\Theta }}{2}-1|~.
\la{slrolPars}
\eeq 
[${\cal V}'$, ${\cal V}''$ denote derivatives with
respect to $\phi_{\Te }$, and $M_{\rm Pl}=2.4\cdot 10^{18}$~GeV].
For moderate values of $\tan^2\fr{\pi g_4R\phi_{\Theta }}{2}$
the slow roll conditions $\ep , |\eta |\ll 1$
can be easily satisfied  by properly suppressed $g_4$.
Therefore, this is a good framework for a natural inflation.
The value $\phi_{\Te }^f$ at which the inflation ends is determined from the conditions $\ep , |\eta |\sim 1$ 
\beq
\tan \fr{\pi g_4R\phi_{\Te }^f}{2}\simeq \fr{\sq{2}}{\pi g_4RM_{\rm Pl}}~,
\la{phif}
\eeq 
(we are considering the interval 
$0\leq g_4R\phi_{\Te }\leq 1$). 
The fulfillment of the slow roll conditions allows to determine analytically the number of ${\rm e}$-foldings during the corresponding time interval
\beq
N(\phi_{\Te }\to \phi_{\Te }^f)=\fr{1}{M_{\rm Pl}^2}
\hs{0.1cm}\int_{\phi_{\Te }^f}^{\phi_{\Te }}
\hs{-0.2cm}\fr{{\cal V}}{{\cal V}'}d\phi_{\Te }
=-\fr{1}{(\pi g_4RM_{\rm Pl})^2}
\ln \l \sin^2 \fr{\pi g_4R\phi_{\Te }}{2}[1+\fr{1}{2}(\pi g_4RM_{\rm Pl})^2]\r 
~.
\la{NTe}
\eeq 
With this expression one can calculate the value $\phi_{\Te}^Q$ which corresponds to the epoch when the present horizon scale crossed outside the inflationary horizon scale. From the present observations we have $N_Q=55-60$
and therefore we need $\pi g_4RM_{\rm Pl}\ll 1$. We will use the latter 
relation for approximating the exact expressions.
Using (\ref{NTe}) we get
\beq
\phi_{\Te }^Q\simeq \fr{2}{\pi g_4R}
\arcsin \l \exp[ -\fr{N_Q}{2}(\pi g_4RM_{\rm Pl})^2] \r \approx 
\fr{0.79}{g_4R}~~~
{\rm with}~~N_Q=60~,RM_{\rm Pl}=10~,g_4=1.4\cdot 10^{-3}.
\la{phiQ}
\eeq 
We see that for this value $\tan^2 \fr{\pi g_4R\phi_{\Te }}{2}$ 
is not small. The slow roll
parameters
\beq
\ep_{Q}\simeq \fr{1}{2N_Q}~,~~~~
\eta_Q\simeq \fr{1}{2N_Q}-\fr{\pi^2}{2}(g_4RM_{\rm Pl})^2
\la{epetaQ}
\eeq
however are small enough and we therefore have the relation 
$3H\dot{\phi }_{\Te }=-{\cal V}'$.

The quadrupole anisotropy of the temperature fluctuations due to the scalar 
perturbations can be calculated according to expression\cite{lid} 
\beq
\l \fr{\de T}{T}\r_{Q-S}=\left. 
\fr{\sq{5}}{60\pi }\fr{{\cal V}^{3/2}}{M_{\rm Pl}^3{\cal V}'}\right|_{\phi_{\Te}^Q}~,
\la{Tgen}
\eeq 
and for our case is given by
\beq
\l \fr{\de T}{T}\r_{Q-S}=-\fr{1}{3\pi^2\sq{10}}\fr{1}{g_4RM_{\rm Pl}}
\l \fr{{\cal V}_0}{M_{\rm Pl}^4}\r^{1/2}
\sinh \l -\fr{N_Q}{2}(\pi g_4RM_{\rm Pl})^2\r \sim 
\fr{N_Q}{8\sq{30}\pi^3}\fr{g_4}{RM_{\rm Pl}}~,
\la{Tfluct}
\eeq 
while the tensor to scalar ratio 
$r=\l \fr{\de T}{T}\r_{Q-T}^2/\l \fr{\de T}{T}\r_{Q-S}^2
\simeq 2.16\pi \l \fr{M_{\rm Pl}{\cal V}'}{\cal V}\r^2$ is
\beq
r\simeq 4.32\pi \ep_Q\simeq \fr{6.8}{N_Q}~.
\la{r}
\eeq 
{}From (\ref{Tfluct}), (\ref{r}) with $N_Q=60$, $g_4=1.4\cdot 10^{-3}$, 
$RM_{\rm Pl}=10$ one obtains the measured value 
$\l \fr{\de T}{T}\r_Q\sim 6\cdot 10^{-6}$. For the same values of the 
parameters one has relatively large $r\simeq 0.11$. This is one of the 
remarkable feature of this inflationary scenario. The planned measurements 
of the Planck satellite could detect such a value of the tensor contribution.
It is interesting to 
note that a quadratic inflaton potential gives a similar relation 
($r\simeq 6.8/N_Q$.), 
although the scenario considered there differs from ours
in various aspects. The spectral index 
$n_s=1+2\eta_Q-6\ep_Q$ for the scenario considered here is
\beq
n_s =1-\fr{2}{N_{Q}}-\pi^2(g_4RM_{\rm Pl})^2 ~.
\la{ns}
\eeq
We see that the spectrum is red-tilted. 
{}For  parameters given in (\ref{phiQ}) we have $n_s\simeq 0.96$, 
which is  compatible with WMAP data \cite{wmap}. 
Combining WMAP and the Ly$\al $
data \cite{Seljak:2004xh} gives the restriction $n_s\stackrel{>}{_\sim }0.96$.
In Table 1 we present the results for several cases satisfying this data.
%
%
\begin{table} \caption{Spectral properties from gauge inflation with 
different values of parameters and
$\l \fr{\de T}{T}\r_Q\sim 6\tm 10^{-6}$.}
 
\label{ginfl} $$\begin{array}{|c|c|c|c|c|c|}
 
\hline 
N_Q & RM_{\rm Pl}  & g_4 & n_s & r   & 10^3\tm \fr{dn_s}{d\ln k}   \\
 
\hline \hline
 
55 & 5 &7.4\cdot 10^{-4} &0.96 &0.12 &-0.67 \\ 
 
\hline

55 & 10 &1.5\cdot 10^{-3} &0.96 &0.12 &-0.78 \\ 

\hline  \hline

 60&5  &7\cdot 10^{-4} &0.97 &0.11 &-0.56 \\ 

\hline

 60&10  &1.4\cdot 10^{-3} &0.96 &0.11 &-0.65 \\ 

\hline

\end{array}$$
 
\end{table}
%
%
%
As we see the tensor to scalar ratio $r$ is significant while the spectral 
index  practically shows no running. 
One can check that for  presented cases
$\phi_{\Te }^Q>M_{\rm Pl}$. However, since $\phi_{\Te }$ corresponds to the gauge field $A_5^1$ one can be sure that 5D gauge invariance and locality will guarantee that there is no undesirable corrections to the inflaton potential. The non local operators, not respecting the shift symmetry 
$\phi_{\Te }\to \phi_{\Te }+\fr{2k}{g_4R}$, are suppressed by a factor 
$e^{-2\pi RM_5}$ \cite{Arkani-Hamed:2003wu}, where $M_5$ is the 5D Planck scale. {}For $R\stackrel{>}{_\sim }\fr{5}{M_5}$ 
the suppression factor is 
$e^{-2\pi RM_5}\stackrel{<}{_\sim } 10^{-13}$. Therefore, all kind of non 
local contributions can be safely ignored.

\section{Modulus field driven inflation}

Now we consider the case in which the inflation is only due to modulus field 
$\phi_M$, assuming that $\phi_{\Te }$ is settled in its minimum
$g_4R\phi_{\Te }=1$. For simplicity we will consider the norm function
$\ka^{-1}{\cal N}=(M^0)^3-M^0(M^1)^2$. Using the constraint 
${\cal N}=\ka^{-2}$, the field $\phi_M$
is not canonically normalized when $\lan \phi_M\ran \neq 0$. 
For parameterisation of the very special manifold we introduce a new variable
$t$ such that
\beq
M^0=\ka^{-1}\cosh^{2/3} \al t~,~~~
M^1=\fr{\phi_M}{\sq{2\pi R}}=\ka^{-1}\fr{\sinh \al t}{\cosh^{1/3} \al t}~,
~~~{\rm with}~~~\al=\fr{\ka }{\sq{2\pi R}}=\fr{1}{M_{\rm Pl}}~.
\la{manif}
\eeq
Then the kinetic term has the form
\beq
\fr{1}{2}g(t)(\pl_{\mu }t)^2=\fr{1}{2}(\pl_{\mu }F)^2~,~~~{\rm with}~~~
\sq{g(t)}dt=dF~,~~g(t)=1+\fr{1}{3}\tanh^2\al t~,
\la{cankin}
\eeq
where $F$ is a canonically normalized field playing the role of the modulus 
inflaton. 
Using the dimensionless variables
\beq
\phi_R=R\phi_M~,~~~~t_P=\fr{1}{M_{\rm Pl}}t~,
\la{lessvar}
\eeq
the derivatives can be written as
$$
\fr{d{\cal V}}{dF}=\fr{1}{M_{\rm Pl}}\fr{1}{\sq{g}}
\fr{\pl {\cal V}}{\pl \phi_R}\fr{\pl \phi_R}{\pl t_P}~,
$$
\beq
\fr{d^2{\cal V}}{dF^2}=\fr{1}{M_{\rm Pl}^2}\fr{1}{g}
\l \fr{\pl^2 {\cal V}}{\pl \phi_R^2}\l \fr{\pl \phi_R}{\pl t_P}\r^2
+\fr{\pl {\cal V}}{\pl \phi_R}\l \fr{\pl^2 \phi_R}{\pl t_P^2}-
\fr{1}{2g}\fr{\pl \phi_R}{\pl t_P}\fr{\pl g}{\pl t_P}\r
\r ~.
\la{derFs}
\eeq
Using these relations we can calculate the slow roll parameters $\ep $,
$\eta $ which allows to determine numerically  the point $t_P^f$ corresponding 
to the end of inflation. The $t_P^Q$
can be determined through the number of e-foldings through the relation
\beq
N_Q=\int_{t_P^f}^{t_P^Q} \sq{\fr{g(t_P)}{2\ep(t_P)}}dt_P~.
\la{NQM}
\eeq
Having determined $t_P^Q$ we can calculate the quantities
$\de T/T$, $n_s$ and $r$.
The selection of $g_4, R$ should be done in such a way as to have 
$\fr{\de T}{T}\approx 6\cdot 10^{-6}$.
Numerical study shows that one can have inflation both for \emph{large} and \emph{small} values of $a\equiv \pi g_4 R M_{Pl}$. For $a\gg 1$ we have 
\beq
\pi g_4R\phi_M\gg 1~,~~~~~ t_P\ll 1~.
\la{condM}
\eeq
This means that the metric ${\cal N}_{IJ}$ is nearly diagonal and certain approximations can be done. Namely, using (\ref{condM}) we obtain for $a\gg 1$
\beq
\l \fr{\de T}{T}\r_Q\sim \fr{N_Q}{16\sq{15}\pi^3}\fr{g_4}{RM_{\rm Pl}}~,
\la{dTMod}
\eeq
\beq
n_s\simeq 1-\fr{2}{N_Q}~,~~~~r\simeq 
\fr{2.16\pi }{(\pi g_4RM_{\rm Pl})^2N_Q^2}~.
\la{nsrMod}
\eeq
In the limit $a\ll 1$, during inflation  we have
\beq
\pi g_4R\phi_M\gg 1~,~~~~~ t_P\gg 1~,
\eeq 
and we obtain the following approximate values
\beq\la{dTMod2}
\l \fr{\de T}{T}\r_Q\sim \fr{N_Q}{48\sq{5}\pi^4}\fr{10}{(RM_{\rm Pl})^2}~,
\eeq
\beq\la{nsrMod2}
n_s\simeq 1-\fr{2}{N_Q}~,~~~~r\sim 
\fr{6.5\pi }{10^2N_Q^2}~.
\eeq
In (\ref{dTMod2}),  (\ref{nsrMod2}) we have taken into account that 
$\pi g_4R\phi_M^Q\sim 10$.
The exact numerical results are summarized in Table 2. They confirm that the 
approximations which led to (\ref{dTMod}), (\ref{nsrMod}), (\ref{dTMod2}) and (\ref{nsrMod2}) work well. 
As we see, the spectrum here is also red tilted. However, the tensor to 
scalar ratio $r$ is strongly suppressed. 


%
%
\begin{table} \caption{Spectral properties from modulus inflation with 
different values of parameters and
$\l \fr{\de T}{T}\r_Q\sim 6\tm 10^{-6}$.}
 
\label{minfl} $$\begin{array}{|c|c|c|c|c|c|}
 
\hline 
N_Q & RM_{\rm Pl}  & g_4 & n_s & r   & 10^3\tm \fr{dn_s}{d\ln k}   \\
 
\hline\hline 
 
60 & 77 &1.3\cdot 10^{-4} &0.96 &1.4\tm 10^{-4} & -1.1\\ 
 
\hline

55 & 100 &2.5\cdot 10^{-2} &0.96 &5\tm 10^{-5} &-0.86 \\ 
 
\hline

60 & 100 &2.15\cdot 10^{-2} &0.96 &5\tm 10^{-5} &-0.75 \\ 

\hline  

 55 & 600  &0.136 &0.96 &0 &-0.54 \\ 

\hline

 60&600  &0.13 &0.97 &0 &-0.54 \\ 

\hline

55 & 3000  & 0.7 &0.96 &0 &-0.64 \\ 

\hline

60 & 3500  & 0.7 &0.97 &0 &-0.54 \\ 

\hline

\end{array}$$
 
\end{table}
%
%
%

\vs{0.5cm}

\section{Discussion}

We presented in the previous sections two different scenarious for inflation in the potential \eqref{Vinfl1} plotted in Fig.1. The first case, inflation in the $M^1=0$ axis, is essentially the gauge inflation model of \cite{Arkani-Hamed:2003wu}. As we pointed out there, successful inflation in this direction requires $a\equiv\pi R g_4 M_{Pl}\ll 1$ and $g_4\ll 1$ (see Table 1). The scenario we called modulus inflation does \emph{not} share these constraints. In fact it is possible to realize modulus inflation for both small and large $a$ (see Table 2), and since $g_4\sim 10^{-2}a^{\frac{1}{2}}$, this scenario allows for a not too suppressed (4D) gauge coupling and relatively large compactification radius. This opens up the possibility to embed the modulus inflation scenario in orbifold GUTs. Note also that for $a\gg 1$ we have $\phi_Q < M_{Pl}$ and therefore quantum gravity corrections should not play a r\^ole. 

Concluding, let us remark that within our analysis we have assumed 
that during inflation the size 
of the extra dimension ($R$) is fixed. In our treatment $R$ is related to 
the lowest component ($e_y^5$) of the radion superfield. Its stabilization 
is needed and may
be realized by one of the mechanisms which have been widely discussed in the 
literature \cite{radst}, \cite{luty2,dudasquiros}. If the extra-dimension is stabilized 
in a way that our inflation scenario is not modified significantly, the 
above analysis should remain valid. However this issue goes beyond the scope of this paper.


\vs{0.5cm}

\hs{-0.7cm}{\bf Acknowledgments}

\vs{0.2cm} 
\hs{-0.7cm}We thank Qaisar Shafi for discussion and interesting comments. 
The research of F.P.C. is supported by Funda\c c\~ ao para a 
Ci\^ encia e a Tecnologia (grant   SFRH/ BD/4973/2001).



\bibliographystyle{unsrt}

\end{document}